\documentclass{article}

\usepackage{arxiv}

\usepackage[utf8]{inputenc} 
\usepackage[T1]{fontenc}    
\usepackage{hyperref}       
\usepackage{url}            
\usepackage{booktabs}       
\usepackage{amsfonts}       
\usepackage{nicefrac}       
\usepackage{microtype}      
\usepackage{lipsum}
\usepackage{amsmath}
\usepackage{colonequals}
\usepackage{natbib}
\setcitestyle{authoryear,open={(},close={)}}
\usepackage{amsthm}
\usepackage{amssymb}
\usepackage{algorithm}
\usepackage[noend]{algpseudocode}
\usepackage{mathtools}
\usepackage{bbm}

\DeclarePairedDelimiter\floor{\lfloor}{\rfloor}
\usepackage{booktabs}
\usepackage{graphicx}
\usepackage{enumitem}

\usepackage{tikz}
\usetikzlibrary{automata}
\usetikzlibrary{positioning}  
\usetikzlibrary{arrows}       
\tikzset{node distance=2.5cm, 
         every state/.style={ 
           semithick,
           fill=gray!10},
         initial text={},     
         double distance=2pt, 
         every edge/.style={  
           draw,
           ->,>=stealth',     
           auto,
           semithick}}
\tikzset{vertex/.style={state, minimum size= 5mm, fill=gray!10, inner sep= 0pt, font=\tiny}}
\tikzset{vertexsmall/.style={state, minimum size= 4mm, fill=gray!10, inner sep= 0pt, font=\tiny}}
\tikzset{base/.style={draw, ->,>=stealth', auto, semithick, font=\tiny, sloped, pos = 0.5, above}}
\tikzset{SE/.style={base, thick}}
\tikzset{SN/.style={base, dashed, thick}}
\tikzset{E/.style={base, color=gray!60, thin}}
\tikzset{FE/.style={base, dotted, color=gray!60, thin}}
\let\epsilon\varepsilon
\newtheorem{theorem}{Theorem}[section]
\newtheorem{definition}[theorem]{Definition}

\newtheorem{lemma}[theorem]{Lemma}
\newtheorem{remark}[theorem]{Remark}
\newtheorem{proposition}[theorem]{Proposition}
\usepackage{subcaption}
\usepackage{xcolor}

\algnewcommand{\IIf}[1]{\State\algorithmicif\ #1\ \algorithmicthen}
\algnewcommand{\EndIIf}{\unskip\ \algorithmicend\ \algorithmicif}

\newcommand*\diff{\mathop{}\!\mathrm{d}}

\providecommand{\keywords}[1]
{
	{\small
  \textbf{\textit{Keywords---}} #1
  }
}

\title{State-Dependent Kernel Selection for Conditional Sampling of Graphs}

\author{
  James ~Scott\\
  Department of Mathematics\\
  Imperial College London\\
  London SW7 2AZ\\
  \texttt{james.scott15@imperial.ac.uk} \\
   \And
 Axel ~Gandy\\
  Department of Mathematics\\
  Imperial College London\\
  London SW7 2AZ\\
  \texttt{axel.gandy@imperial.ac.uk} \\
}

\begin{document}
\maketitle
\bibliographystyle{amstat}

\begin{abstract}
This paper introduces new efficient algorithms for two problems: sampling conditional on vertex degrees in unweighted graphs, and sampling conditional on vertex strengths in weighted graphs. 
The algorithms can sample conditional on the presence or absence of an arbitrary number of edges.
 The resulting conditional distributions provide the basis for exact tests. 
Existing samplers based on MCMC or sequential importance sampling are generally not scalable; their efficiency degrades in sparse graphs. MCMC methods usually require explicit computation of a Markov basis to navigate the complex state space; this is computationally intensive even for small graphs.
We use state-dependent kernel selection to develop new MCMC samplers. These do not require a Markov basis, and are efficient both in sparse and dense graphs. The key idea is to intelligently select a Markov kernel on the basis of the current state of the chain. We apply our methods to testing hypotheses on a real network and contingency table. The algorithms appear orders of magnitude more efficient than existing methods in the test cases considered.
\end{abstract}


\section{Introduction}
\label{1}

Inference on graphs conditional on vertex level data arises in many diverse disciplines including network science, psychometrics, community ecology and categorical data analysis. Two difficult, yet important problems involve sampling from the set of graphs (weighted graphs) conditional on prescribed vertex degrees (strengths). In many settings, researchers additionally need to condition on an arbitrary set of known edges/non-edges in the graph.

The resulting conditional distributions are used to perform hypothesis tests.
Approximate tests rely on distributional approximations to the null. These approximations can be arbitrarily poor in sparse graphs, and so alternative approximations, based on sampling, become particularly important. 
Unfortunately, most proposed samplers are inefficient in sparse graphs.

Additionally, if the null distribution conditions on known edges/non-edges, it is difficult to construct a connected Markov chain on the relevant state space. Existing methods either specialise to particular patterns of fixed edges, or in the general case, use techniques from computational algebra to compute a Markov basis. These methods are computationally intensive and are impractical for graphs with more than a few vertices.

We propose a new class of MCMC methods that use \textit{state-dependent mixing} of Markov kernels. This technique allows us to construct samplers that require little tuning to the problem at hand, and do not require computation of a Markov basis. The samplers are irreducible in the face of arbitrary patterns of fixed edges/non-edges.  They are efficient both in extremely sparse and dense graphs, and appear orders of magnitude more efficient than existing methods in the test cases we consider.

The first focus of this paper is on uniformly sampling unweighted graphs conditional on prescribed vertex degrees. In the context of hypothesis testing, the vertex degrees are often sufficient statistics for nuisance parameters in the null distribution; in such applications, conditioning allows researchers to perform exact tests. 

Consider social network analysis. A social network equipped with a dichotomous relation can be expressed as a simple digraph. Vertices represent actors, with edges representing the applicability of the relation between actors. Frequently researchers are interested in testing the presence of reciprocity in the network; defined loosely as a preference for mutual dyads in the corresponding digraph.

\citet{holland_leinghardt_1981} introduce an exponential family model under which the UMPU test for reciprocity conditions on the observed degree sequences. In this case, conditioning serves to remove unknown parameters from the null, and the resulting distribution is then uniform on the reference set. 

The complex interactions that result from conditioning render analytic analysis of the null distribution difficult or impossible. Efforts have been made to develop recursive formulas to enumerate all graphs in the reference set \citep[see][]{wasserman_1994}, however these are impractical for even moderately sized graphs.

If we can sample graphs (nearly) uniformly, then we can approximate the null distribution of an arbitrary test statistic. Thus, the literature has focused on simulation, whose methods can broadly be divided into two camps; Markov Chain Monte Carlo (MCMC) \citep{rao_1996,roberts_2000,milo_2002,mcdonald_2007,verhelst_2008} and sequential importance sampling (SIS) \citep{snijders_1991,zhang_chen_2013,chen_2005,bayati_2010}.

A digraph can be represented by its adjacency matrix. Thus, the sampling problem is equivalent to uniformly sampling zero-one tables with given margins and a structurally zero diagonal. Sampling zero-one tables with given margins is applied in community ecology to test for patterns in co-occurrence tables, and in psychometrics to test the Rasch hypothesis \citep[see][]{gustafsson_1980}. Thus, there exists a substantial parallel literature along these lines. 

Most MCMC algorithms proposed for sampling graphs are adaptations of methods proposed for zero-one tables. Typically, they use  a combination of `switch' moves \citep{ryser_1963} and additional moves to maintain irreducibility in the face of structural zeros. \citet{rao_1996} and \citet{mcdonald_2007} consider `compact alternating hexagon' and `hexad' updates respectively. Most proposed methods suffer from poor mixing in unbalanced matrices, rendering them impractical for moderate to large graphs. Additionally, they are not extensible to arbitrary patterns of fixed edges and non-edges.

SIS builds the graph sequentially, at each iteration choosing a candidate edge with probability proportional to the vertex degrees. Early methods for this application include \citep{snijders_1991, chen_2005}. Most of these samplers get stuck, and the probability of restarting approaches 1 as the degree sequences grow. \citet{bezakova_2012} provide examples where such algorithms are slow. More recent methods avoid the issue of restarting and often come with better theoretical guarantees \citep{ bayati_2010,blitzstein_diaconis_2011,zhang_chen_2013}. 

The second focus of this paper is on sampling integer-weighted graphs given vertex strengths. This can be used to conduct network tomography in the case of a star network topology. However, the motivating application is approximating the null distribution for evaluating exact tests on two-way contingency tables. This is a classical problem in statistics which is important because standard asymptotics justifying approximate tests (notably Pearson’s $\chi^2$ test of independence) do not hold for tables with cells with low expected frequencies \citep[see][]{agresti2003}. 

In conditional tests of independence one is interested in the hypergeometric distribution on all tables with given margins. This corresponds to sampling integer-weighted bipartite graphs conditional on vertex strengths. \citet{diaconis_sturmfels_1998} proposed a simple ‘switch’ Markov chain to sample from such tables. We describe this in more detail in Section \ref{sdks}. It suffers slow mixing in sparse tables. 

\citet{diaconis_sturmfels_1998} also proposed an algebraic algorithm to construct a connected Markov chain in the context of incomplete tables. Other MCMC methods proposed to sample incomplete tables also rely on computing a Markov basis \citep{aoki_2005,rapallo_2006}. The computational cost of computing a Markov basis is exponential in the size of the table. Additionally, the computation is example specific; i.e. a new basis must be computed for each pattern of structural zeros considered.

\citet{chen_2005} introduced the first SIS method for uniform sampling of contingency tables with given marginals. \citet{chen_2007} extended this to incomplete tables. \citet{eisinger_2017} develop a sampler with improved efficiency, particularly in sparse graphs. We will compare our sampler to SIS approaches in our applications.

An R-package implementing the new algorithms is available at \href{https://github.com/jscott6/cgsampr}{https://github.com/jscott6/cgsampr}.

\section{Notation and Background}
\label{prelim}

A graph $G \colonequals (V, E)$ is a pair with $V$ being the vertex set and $E$ the set of edges. Throughout we assume an unweighted graph has no multiple edges, and note that the other case can be dealt with as an integer-weighted graph. If the context requires clarification, we use $V(G)$ and $E(G)$ to denote the sets belonging to $G$. We denote an edge from vertex $u$ to vertex $v$ by $uv$. If the graph is undirected, $uv$ is equivalent to $vu$. An integer-weighted graph is a triple $G \colonequals (V, E, c)$. The function $c\colon V \times V \rightarrow \mathbb{N}_0$ assigns a positive integer valued weight to each $uv \in E(G)$, and $0$ to each $uv \notin E$. 

The in- and out-degrees of a vertex are the number of edges to and from the vertex respectively. The in- and out-strengths of a vertex of a weighted graph are the sum of the weights of edges to and from the vertex respectively. If the graph is undirected, there is no distinction between in and out, so we simply use the terms degree and strength of a vertex.

Two undirected graphs with the same vertex set have the same degree sequence if every vertex has the same degree in both graphs. A similar terminology is used for directed graphs, where both the in- and out-degrees have to be equal for every vertex. We use analogous definitions for equivalent vertex strength sequences.

\section{State-Dependent Kernel Selection}
\label{sdks}

Consider the following problem. Let $r \colonequals (r_1,...,r_I)$ and $c \colonequals (c_1,...,c_J)$ be non-negative integer vectors, and let $\mathcal{X}$ denote the set of all $I\times J$ non-negative integer matrices such that the row and column marginals equal $r$ and $c$ respectively. Assume $\mathcal{X}$ is non-empty. The task is to construct a Markov chain ergodic with respect to the uniform distribution on $\mathcal{X}$.

\citet{diaconis_sturmfels_1998} describe a simple Markov chain for this purpose. Given $X_n = x$, pick a pair of rows and a pair of columns uniformly at random. The walk proceeds by sampling from the conditional distribution of the delineated subtable given all other entries. An update takes the form

$$
\begin{matrix}
+\Delta & -\Delta \\
-\Delta & +\Delta 
\end{matrix}
$$
for $\Delta$ sampled uniformly from the admissible range: integers which do not induce negative values in the subtable.

A Markov chain on $\mathcal{X}$ is completely characterized by its (Markov transition) kernel $K$, a regular conditional distribution, where $K(x, A)$ represents the probability that the next state of the chain is in $A$ measurable given that the current state is $x$. 

In this example, the kernel $Q$ of the chain can be viewed as randomly selecting from a set of other kernels. Indeed, let $\mathcal{Z}$ be the collection of indices of all $2\times 2$ sub-arrays of $I \times J$ tables. The Gibbs update along each $z \in \mathcal{Z}$ defines a kernel $K_z$ on $(\mathcal{X},\mathcal{B})$. We define a \textit{scan order} as a method of choosing a particular kernel from this collection at each iteration of the chain. The aforementioned chain is an example of a \textit{random scan} procedure; corresponding to mixing the kernels $\{K_z\}$. The kernel of the chain is then $Q \colonequals \sum_{z\in \mathcal{Z}} K_z /|\mathcal{Z}|$, where $|\mathcal{Z}|$ represents the cardinality of $\mathcal{Z}$.

The chain suffers poor mixing in sparse matrices as $\Delta$ is often degenerate at 0. State-dependent kernel selection asks whether we can improve mixing by allowing the scan order to depend on the current state of the Markov chain, whilst maintaining ergodicity with respect to the target distribution. 

In the Diaconis-Sturmfels chain a useful strategy might attach density only on the subset of $\mathcal{Z}$ for which the range of $\Delta$ would be non-zero. However, this will not generally maintain invariance with respect to $\pi$. As we will see, this can be overcome by either carefully constructing the scan order, or by modifying the kernels $K_z$ themselves.

More generally, suppose we have a collection of kernels $K \colonequals \{ K_z \colon z \in \mathcal{Z}\}$ on a measurable space $(\mathcal{X},\mathcal{B})$. Equip the index set $\mathcal{Z}$ with a sigma-algebra $\mathcal{F}$. One defines a conditional distribution $\mathbb{V}$, on $\mathcal{F}$ given the current state of the chain. For each $x\in\mathcal{X}$, the measure $\mathbb{V}_x$ denotes the law of a random variable representing the kernel from which to sample next. 
We will assume throughout that the map $z \rightarrow K_z(x,A)$ is $\mathcal{F}$-measurable for each $x \in \mathcal{X}$ and $A\in\mathcal{B}$.

The kernel of the state-dependent Markov chain $Q$ on $(\mathcal{X},\mathcal{B})$ is defined by the integrals

\begin{equation} \label{decomp}
Q(x,\cdot) \colonequals \int K_z(x,\cdot) \mathbb{V}_x(\diff z)\text{ for all } x\in\mathcal{X},
\end{equation}
which are measures in the second argument. Suppose a Markov chain defined by $Q$ has some current state $x$. The chain samples some kernel $K_z$ according to the measure $\mathbb{V}_x$, and proceeds to sample the next state of the chain from the measure $K_z(x,\cdot)$.

\subsection{Decomposing a Kernel}
\label{sec:DecKernel}

Any Markov kernel can be represented as a weighted average of a collection of kernels.  Definition \ref{decompositiondef} helps make this precise.

\begin{definition} \label{decompositiondef}
$(K,\mathbb{V})$ is a \textit{decomposition} of a kernel $Q$ on $(\mathcal{X},\mathcal{B})$ if \eqref{decomp} holds.
\end{definition}

Definition \ref{decompositiondef} implies that every kernel $Q$ has an `identity' decomposition, given by $(\{K_1\},\mathbb{V})$ with $K_1=Q$ and $\mathbb{V}_x(\{1\})=1$ for all $x \in \mathcal{X}$. The decomposition of a kernel is typically not unique.

\begin{definition} \label{symmetrickernel}
Let $(K,\mathbb{V})$ be a decomposition of $Q$. Suppose $\mathbb{V}_x$ is dominated by a $\sigma$-finite measure $\mu$ for all $x \in \mathcal{X}$. Then, the decomposition is symmetric if there exist densities $\{f_x\}$ (each with respect to $\mu$) such that for each $z$ and $x$, $f_x(z) = f_y(z)$ for $K_z(x,\cdot)$-almost every $y$.
\end{definition}

Suppose $Q$ has a symmetric decomposition $(K,\mathbb{V})$. Definition \ref{symmetrickernel} implies that, at any given iteration of the chain, once $K_z$ has been sampled, the density of sampling $K_z$ under almost every possible new state of the chain is unchanged. Any state-independent kernel selection is symmetric (for example, random scan). The `identity' decomposition of any kernel is symmetric. Definition \ref{symmetrickernel} generalises this concept to state-dependent mixing.

\begin{figure}[tbp]
  \centering
  \begin{subfigure}[b]{0.4\linewidth}
  \centering
  \begin{tikzpicture}
    \node[state] (1) at (1.5,2.6) {$1$};
    \node[state] (2) at (0,0) {$2$};
    \node[state] (3) at (3,0) {$3$};
    \draw (1) edge[bend left = 15, dashed] node {\tt $1$} (2)
              edge[bend left = 15] node {\tt $1$} (3)
              edge[loop above, densely dotted] node {\tt $1$} (1)
          (2) edge[bend left = 15, dashed] node {\tt $1$} (1)
              edge[bend left = 15, densely dotted] node {\tt $1$} (3)
              edge[loop left] node {\tt $1$} (2)
          (3) edge[bend left = 15] node {\tt $1$} (1)
              edge[bend left = 15, densely dotted] node {\tt $1$} (2)
              edge[loop right, dashed] node {\tt $1$} (2);
  \end{tikzpicture}
  \end{subfigure}
  \begin{subfigure}[b]{0.4\linewidth}
  \centering
  \begin{tikzpicture}
    \node[state] (1) at (1.5,2.6) {$1$};
    \node[state] (2) at (0,0) {$2$};
    \node[state] (3) at (3,0) {$3$};
    \draw (1) edge[bend left = 15] node {\tt $.5$} (2)
              edge[bend left = 15] node {\tt $.5$} (3)
          (2) edge[bend left = 15] node {\tt $.5$} (1)
              edge[bend left = 15] node {\tt $.5$} (3)
          (3) edge[bend left = 15] node {\tt $.5$} (1)
              edge[bend left = 15] node {\tt $.5$} (2);
  \end{tikzpicture}
  \end{subfigure}
  \caption{\small{Kernels used as examples of a decomposition in Section \ref{sec:DecKernel}. Left: $K_1$ (dotted), $K_2$ (solid), $K_3$ (dashed). Right: Kernel $Q$} }
  \label{threestate}
\end{figure}
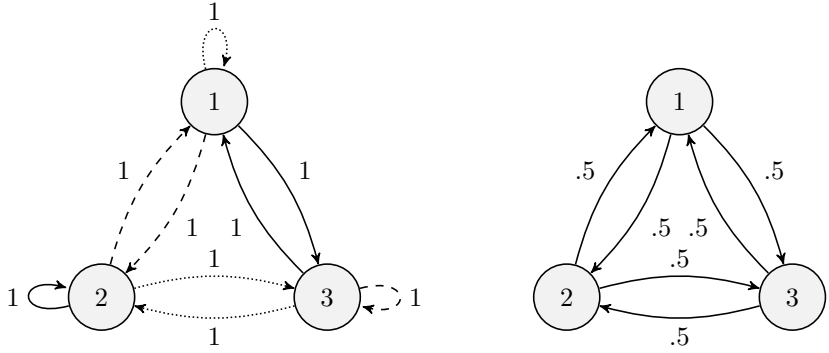

As an example, consider a three-state state space, as depicted in Figure \ref{threestate}. The left figure defines three kernels on this space. Let $K_1$, $K_2$ and $K_3$ be defined by the dotted, solid and dashed transitions in the figure respectively. A naive chain might pick from these kernels randomly, irrespective of the current state. If the chain is in state $i$, the kernel $K_i$ cannot change the state. A better strategy might pick randomly from the other two kernels, guaranteeing that the chain moves to a new state. This state-dependent strategy is easily seen to lead to a symmetric decomposition.

\begin{lemma} \label{symmetrictheorem}
A kernel $Q$ is reversible with respect to a distribution $\pi$ if and only if it has a symmetric decomposition $(K,\mathbb{V})$ where every $K_z \in K$ is reversible with respect to $\pi$.

\end{lemma}

If both $\mathcal{X}$ and $\mathcal{Z}$ are countable, then it is easy and instructive to illustrate detailed balance when the conditions of Lemma \ref{symmetrictheorem} hold. In particular, suppose $(K,\mathbb{V})$ is symmetric and each $K_z$ is reversible with respect to the uniform distribution on $\mathcal{X}$. Then

$$
Q(x,y) = \sum_{z\in \mathcal{Z}}\mathbb{V}_z(x)K_z(x,y) = \sum_{z\in \mathcal{Z}}\mathbb{V}_z(y)K_z(y,z) = Q(y,x).
$$
We have used the fact that $\mathbb{V}_x(z) = \mathbb{V}_y(z)$ whenever $K_z(x,y) > 0$, and that $K_z(x,y) = K_z(y,x)$. 

Recall the three state Markov chain, shown in Figure \ref{threestate}. Each of the three kernels is reversible with respect to the uniform distribution. Additionally, the decomposition under the state-dependent strategy previously suggested is symmetric. Lemma \ref{symmetrictheorem} thus implies the overall chain (whose transitions are shown in Figure \ref{threestate}, right) also conserves the uniform distribution.

Lemma \ref{symmetrictheorem} implies we can transform the problem of establishing reversibility of $Q$ with respect to a distribution into a problem of finding a decomposition $(K, \mathbb{V})$ of $Q$ satisfying particular conditions. This will be our strategy in Section \ref{unweighted}.

\subsection{Kernel Selection as an Auxiliary Variable}
\label{auxvar}
We cannot usually expect a kernel decomposition $(K,\mathbb{V})$ to be symmetric. We now outline a more general sampling strategy. The kernel $Q$ can be interpreted as defining a Markov chain $(Z_n, X_n)_{n\geq 0}$ on the extended space $\mathcal{Z} \times \mathcal{X}$. Let the current state of this chain be $y := (z, x)$. If $y'$ represents the next state then

\begin{equation} \label{enlarged_chain}
\Pr (y' \in A_1 \times A_2 \mid y) = \int_{A_1} K_z(x,A_2) \mathbb{V}_x(\diff z)
\end{equation}
for $A_1$ and $A_2$ measurable. If the chain defined by \eqref{enlarged_chain} is ergodic with respect to the joint distribution $\mathbb{V} \otimes \pi$, then the sub-chain $(X_n)$ with kernel $Q$ is ergodic with respect to its marginal distribution $\pi$. This suggests a two stage sampler, a step of which is shown below.

\begin{enumerate}
  \item Sample $Z_n \sim \mathbb{V}_{X_{n-1}}$.
  \item Sample $X_n \sim \mathbb{V} \otimes \pi(\cdot \mid Z_n, T_{Z_n}(X_{n-1}))$.
\end{enumerate}
Each $T_{z}$ is a statistic on $\mathcal{X}$. Both steps are Gibbs kernels reversible with respect to the joint distribution $\mathbb{V} \otimes \pi$, and so their composition also conserves the joint. We use this strategy in Section \ref{sec:weighted}.

\section{Sampling Unweighted Graphs}
\label{unweighted}

Let $G_0$ be a given directed or undirected graph with a finite vertex set $V$. 
Let $\mathcal{F}$ be a subset of possible edges of a graph with vertex set $V$.
Let $\mathcal{G}$ be the set of all graphs $G$ with the same vertex set and degree sequence as $G_0$, and additionally satisfying $E(G)\cap \mathcal{F}=E(G_0)\cap \mathcal{F}$.
Our goal is to sample from the uniform distribution on $\mathcal{G}$.

Intuitively, the set $\mathcal{F}$ represents edges known \textit{by design} to be present or absent. Given vertices $u$ and $v$, if $uv$ belongs to $\mathcal{F}$ then $uv$ is either present in all graphs in $\mathcal{G}$, or in none. We stress \textit{by design} because the constraints imposed by the degree sequence and $\mathcal{F}$ may imply that further edges are present or absent in all graphs of $\mathcal{G}$. We call this set $\tilde{\mathcal{F}}$ the set of known edges,  and formally define it as 
$$
\tilde{\mathcal{F}}=\{\text{possible edges }uv: uv\in G_0\Leftrightarrow (uv\in G \text{ for all } G\in \mathcal{G})\}
$$
We show in Section \ref{findfixed} a method of obtaining $\tilde{\mathcal{F}}$.

For every graph $G \in \mathcal{G}$, we will extensively use two `neighborhood' sets associated to each vertex $u$. The set $N_G(u)$ are the in-neighbors of $u$, excluding any vertex $v$ for which the edge $vu$ is known. $M_G(u)$ is the set of all vertices $v$ which are not out-neighbors of $u$, and for which the absence of $uv$ is not known. These are defined as
$$ 
N_G(u) \colonequals \{ v \in V  \colon vu \in E(G), vu \notin \tilde{\mathcal{F}} \}, 
$$
$$
M_G(u) \colonequals \{ v \in V \colon uv \notin E(G), uv \notin \tilde{\mathcal{F}} \}. 
$$

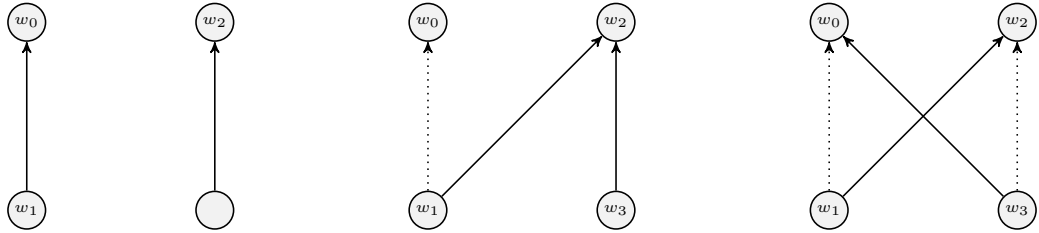
\begin{figure}[h]
\centering
\begin{subfigure}[b]{0.3\linewidth}
\centering
\begin{tikzpicture}
  \node[vertex] (1) at (0,2.5) {$w_0$};
  \node[vertex] (2) at (2.5,2.5) {$w_2$};
  \node[vertex] (3) at (0,0) {$w_1$};
  \node[vertex] (4) at (2.5,0) {};
  \draw (3) edge node {} (1)
  	(4) edge node {} (2);
\end{tikzpicture}
\end{subfigure}
\hfill
\begin{subfigure}[b]{0.3\linewidth}
\centering
\begin{tikzpicture}
  \node[vertex] (1) at (0,2.5) {$w_0$};
  \node[vertex] (2) at (2.5,2.5) {$w_2$};
  \node[vertex] (3) at (0,0) {$w_1$};
  \node[vertex] (4) at (2.5,0) {$w_3$};
  \draw (3) edge[dotted] node {} (1)
  			edge node {} (2)
  		(4) edge node {} (2);
\end{tikzpicture}
\end{subfigure}
\hfill
\begin{subfigure}[b]{0.3\linewidth}
\centering
\begin{tikzpicture}
  \node[vertex] (1) at (0,2.5) {$w_0$};
  \node[vertex] (2) at (2.5,2.5) {$w_2$};
  \node[vertex] (3) at (0,0) {$w_1$};
  \node[vertex] (4) at (2.5,0) {$w_3$};
  \draw (3) edge[dotted] node {} (1)
  			edge node {} (2)
  		(4) edge[dotted] node {} (2)
        	edge node {} (1);
\end{tikzpicture}
\end{subfigure}
\caption{\small{One iteration of Algorithm \ref{binarysampler} with two iterations in the while loop. Left and center show graphs and quantities immediately prior to the first and second edge swaps respectively. Right shows the returned graph. This move corresponds to the well known `switch' step originating from \citet{ryser_1963}.}}
\label{unweightedsamplingstep}
\end{figure}

Here is a sketch of one iteration of the scheme. An initial vertex $W_0$ is sampled uniformly from the set of all vertices $v$ for which $N_G(v)$ is non-empty, and set $n=0$. Sample  $W_1$ uniformly from $N_G(W_0)$, then sample $W_2$ uniformly from $M_G(W_1)$. Replace $W_1W_0$ in $E(G)$ with $W_1W_2$. Letting $n = n + 2$, iterate this procedure, however in each subsequent step $W_{n+1}$ cannot be $W_{n-1}$; this prevents the sampler adding the edge $W_{n+1}W_{n+2}$, and removing it in the next iteration, and should improve state space exploration. Iterate until $W_n$ is $W_0$, at which point all degrees have been maintained. Algorithm \ref{binarysampler} makes this precise. Figure \ref{unweightedsamplingstep} shows a straightforward example of one step of the sampler.

\begin{algorithm}[tbp]
\small
\caption{\small{One Iteration of the Unweighted Graph Sampler (UGS)}}
\label{binarysampler}
\begin{algorithmic}[1]
\Require $G$, $\tilde{\mathcal{F}}$
\State $W_{-1} \gets *$
\State $W_0 \sim U(\{ v \in V\colon N_G(v) \neq \emptyset \})$
\State $n \gets 0$
\Repeat
\State $W_{n+1} \sim U(N_G(W_n) \setminus \{W_{n-1}\})$
\State $W_{n+2} \sim U(M_G(W_{n+1}))$
\State $E(G) \gets E(G)\setminus \{ W_{n+1}W_n\}$
\State $E(G) \gets E(G) \cup \{ W_{n+1}W_{n+2}\}$
\State $n \gets n+2$
\Until {$W_n=W_0$}
\end{algorithmic}
\Return $G$
\end{algorithm}

\subsection{Properties of Algorithm \ref{binarysampler}}
\label{sec:prop_binsampler}
Let $a^r$ denote the reverse of a finite sequence $a$. Given a graph, let $a_1b_1 \leftrightarrow a_2b_2$ denote the operation of replacing the edge $a_1b_1$ with the edge $a_2b_2$. We will refer to this operation as an (edge) \textit{swap}. We call $a_1b_1 \leftrightarrow a_2b_2$ \textit{viable} if and only if $a_1b_1$ is an edge, $a_2b_2$ is not an edge and both $a_1b_1$ and $a_2b_2$ are not in $\mathcal{F}$.

A single iteration of Algorithm \ref{binarysampler} samples a (random) sequence of vertices $W$. Proposition \ref{finitez} implies that this sequence will be finite, so that $W$ takes the form $w_0w_1...w_kw_0$ for some $k$ odd. Let $\mathcal{W}$ be the collection of sequences taking this form.

\begin{proposition} \label{finitez}
For any input graph $G \in \mathcal{G}$ and any $\mathcal{F}$, the expected length of the vertex sequence $W$ (formed by Algorithm \ref{binarysampler}) is finite.
\end{proposition}

Let two sequences be \textit{equivalent} if and only if they are either identical or they are each others' reverse. We let $\mathcal{Z}$ be the quotient set of $\mathcal{W}$ by this equivalence relation. 

We will associate each equivalence class $z \in \mathcal{Z}$ with a kernel on $\mathcal{G}$. Fix any $z$ and let $w$ be a representative of $z$. Consider the following Markov chain on $\mathcal{G}$. From the current state, attempt to iteratively perform the swaps $w_1w_0 \leftrightarrow w_1w_2$, $w_3w_2 \leftrightarrow w_3w_4$, ... ,  $w_kw_{k-1} \leftrightarrow w_kw_0$ to obtain the next state. We say this move is \textit{viable} if and only if all of the swaps are viable when applied iteratively. We refer to this sequence of swaps as the \textit{swaps corresponding to $w$}. If the swaps are not viable, attempt the swaps corresponding to $w^r$; i.e. $w_kw_0 \leftrightarrow w_kw_{k-1}$, ..., $w_1w_2 \leftrightarrow w_1w_0$. If neither swap sequence is viable, then the next state of the chain is unchanged. We define $K_z$ as the kernel of this chain. Remark \ref{welldefined} implies that $K_z$ is well-defined; specifically, the definition is independent of the chosen representative of $z$.

\begin{remark} \label{welldefined}
If the sequences $w$ and $w^r$ are distinct and the swaps corresponding to $w$ are viable, then the swaps corresponding to $w^r$ are not viable.
\end{remark}

Let $K$ be the collection of these kernels. The conditional distribution $\mathbb{V}$ on $\mathcal{Z}$ is defined implicitly by the law of $W$. Formally, the sampler selects a kernel $K_z$ by sampling a vertex sequence $w \in z$. $K_z$ would also be selected if $w^r$ were sampled.
Lemma \ref{symdecomp} implies that $Q$ is reversible with respect to the uniform distribution on $\mathcal{G}$. 

\begin{lemma} \label{symdecomp}
$(K,\mathbb{V})$ is a symmetric decomposition of $Q$, and each $K_{z} \in K$ is reversible with respect to the uniform distribution on $\mathcal{G}$.
\end{lemma}

Proposition \ref{allstates} holds by additionally showing the chain is connected.

\begin{proposition} \label{allstates}
The Markov chain is ergodic with respect to the uniform distribution on $\mathcal{G}$.
\end{proposition}

\subsection{Identifying all Known Edges/Non-Edges}
\label{findfixed}

We show how to determine all structurally fixed edges/non-edges prior to sampling; or in other words, how to determine $\tilde{\mathcal{F}}$ from $\mathcal{F}$ and the degree sequence. Our method makes use of auxiliary graphs, which we now define.

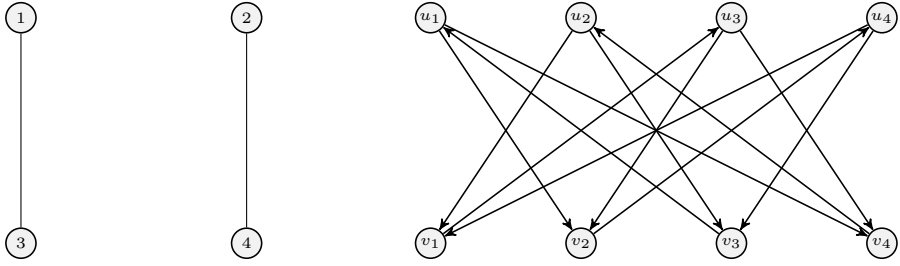
\begin{figure}[tbp]
\centering
\begin{subfigure}[b]{0.3\linewidth}
\centering
\begin{tikzpicture}
  \node[vertexsmall] (1) at (0,3) {$1$};
  \node[vertexsmall] (2) at (3,3) {$2$};
  \node[vertexsmall] (3) at (0,0) {$3$};
  \node[vertexsmall] (4) at (3,0) {$4$};
  \draw (3) -- (1)
  		(4) -- (2);
\end{tikzpicture}
\end{subfigure}
\begin{subfigure}[b]{0.6\linewidth}
\centering
\begin{tikzpicture}
  \node[vertexsmall] (u1) at (0,3) {$u_1$};
  \node[vertexsmall] (u2) at (2,3) {$u_2$};
  \node[vertexsmall] (u3) at (4,3) {$u_3$};
  \node[vertexsmall] (u4) at (6,3) {$u_4$};
  \node[vertexsmall] (v1) at (0,0) {$v_1$};
  \node[vertexsmall] (v2) at (2,0) {$v_2$};
  \node[vertexsmall] (v3) at (4,0) {$v_3$};
  \node[vertexsmall] (v4) at (6,0) {$v_4$};
  \draw (u1) edge node {} (v2)
  			 edge node {} (v4)
        (u2) edge node {} (v1)
        	 edge node {} (v3)
        (u3) edge node {} (v2)
             edge node {} (v4)
        (u4) edge node {} (v1)
        	 edge node {} (v3)
        (v1) edge node {} (u3)
        (v2) edge node {} (u4)
        (v3) edge node {} (u1)
        (v4) edge node {} (u2);
\end{tikzpicture}
\end{subfigure}
\caption{\small{Left: an undirected graph. Right: Auxiliary graph constructed from it, as outlined in Section \ref{findfixed}.}}
\label{exampleauxiliary}
\end{figure}

Given any graph $G \in \mathcal{G}$ we construct an auxiliary bipartite digraph $B \colonequals (U,V,E)$ as follows. Let $U = \{u_i\}$ and $V = \{v_i\}$ for $i = 1,...,n$. Fix any vertex $i$ and any vertex $j$ in $V(G)$. We will add an edge to $B$ if and only if $ij$ is not in $\mathcal{F}$. If additionally $ij$ is in $E(G)$, add $v_ju_i$ to $E(B)$, otherwise add $u_iv_j$. Figure \ref{exampleauxiliary} shows an example of one such graph.

Let $B_\mathcal{G}$ denote the collection of all graphs generated this way from the set $\mathcal{G}$. Proposition \ref{stronglyconnected} shows that we can identify all known edges/non-edges prior to sampling by identifying all strongly connected components of any graph in $B_{\mathcal{G}}$. This can be done using a depth-first search on $B_G$, followed by another depth-first search on the transposed graph. The running time of this preprocessing procedure is $\Theta(2n+n^2)$ \citep[][chap. 22]{cormen_2009}.

\begin{proposition} \label{stronglyconnected}
Fix any graph $B \in B_\mathcal{G}$, and partition its vertex set into strongly connected components $S_1,...,S_K$. The vertex pair $ij$ belongs to $\tilde{\mathcal{F}}$ if and only if either there is no edge incident to $u_i \in S_k$ and $v_j \in S_l$, or if $k \neq l$.
\end{proposition}

\section{Sampling Weighted Graphs}
\label{sec:weighted}

Let $G_0$ be a given integer-weighted directed or undirected graph with a finite vertex set $V$. 
Let $\mathcal{F}$ be a subset of possible edges of a graph with vertex set $V$.
Let $\mathcal{G}$ be the set of all graphs $G$ with the same vertex set and strength sequence as $G_0$, and additionally satisfying $c_G(uv) = c_{G_0}(uv)$ for all $uv \in \mathcal{F}$.
Our goal is to sample from the uniform distribution on $\mathcal{G}$.

Define $\mathcal{Z}$ as in Section \ref{sec:prop_binsampler}. Fix $z \in \mathcal{Z}$ and let $w$ be a vertex sequence belonging to $z$. Let $n_w(uv)$ denote the difference between the number of occurrences of the possible edge $uv$ in $w_1w_0$, $w_3w_2$,...,$w_kw_{k-1}$ and the number of occurrences in $w_1w_2$, $w_3w_4$,...,$w_kw_0$.

We will define $K_z$ as the kernel of the following Markov chain. From the current state $G$, the chain can only move to graphs whose edge strengths take the value $c_G(uv) + n_w(uv)\Delta$ for all $uv$, and for some $\Delta$ in its admissible range: integers for which the resulting graph is in $\mathcal{G}$. 
We will denote the range of $\Delta$ by $[\Delta_{low}, \Delta_{up}]$. 
Let $G_{\Delta^*}$ be the graph obtained by sampling $\Delta = \Delta^*$.
The next state of the chain is determined by sampling $\Delta$ proportional to $\mathbb{V}_{G_{\Delta^*}}(Z_n)$ for each $\Delta^* \in [\Delta_{low}, \Delta_{up}]$, where $\mathbb{V}$ is the (as yet, undefined) law representing our kernel selection strategy.

Observe that there exists a statistic $T_z$ satisfying

$$
T_z^{-1} \circ T_z(G) = \{G_{\Delta^*}: \Delta^* \in [\Delta_{low}, \Delta_{up}]\}.
$$

Thus, sampling $\Delta$ as suggested is the same as the second stage of the sampling scheme presented in Section \ref{auxvar}, and the overall chain defined by $Q$ will will be reversible with respect to the uniform distribution on $\mathcal{G}$.

\subsection{Kernel Selection Strategy}
\label{kss}

$\Delta$ will be degenerate at $0$ if there exists $uv$ and $u'v'$ such that $n_w(uv)$ is positive, $n_w(u'v')$ is negative, and $c_G(uv) = c_G(u'v') = 0$. As the sparsity of $G$ increases, the proportion of kernels in the collection $K$ which can move the chain to a new state decreases. Thus, any state-independent strategy will suffer a degradation in performance. Our state-dependent strategy avoids this.

We alternately samples from two sets associated with each vertex $u$. We redefine $N_G(u)$ and $M_G(u)$ as

$$ 
N_G(u) \colonequals \{ v \in V  \colon vu \in E(G), vu \notin \mathcal{F} \}, 
$$
$$
M_G(u) \colonequals \{ v \in V \colon uv \notin \mathcal{F} \}. 
$$

The kernel sampling procedure is as follows. Let $W_{-1} = *$ and sample $W_0 \sim U(\{v \in V\colon N_G(v) \neq \emptyset \})$. Letting $n=0$, repeat the following until termination. 

\begin{figure*}[tbp]
\centering
\begin{subfigure}[b]{0.3\linewidth}
\centering
\begin{tikzpicture}
	\foreach \phi in {0,...,3}{
    	\node[vertex] (\phi) at (360/8 * \phi:2cm) {$w_\phi$};
     }
     \foreach \phi in {4,...,7}{
    	\node[vertex] (\phi) at (360/8 * \phi:2cm) {};
     }
     \draw (1) edge[SE] node {$+\Delta$} (0)
  			   edge[SN] node {$-\Delta$} (2)
               edge[E] node {} (6)
  		   (3) edge[SE] node {$+\Delta$} (2)
               edge[SE] node {$-\Delta$} (0)
               edge[E] node {} (6)
           (5) edge[E] node {} (4)
               edge[E] node {} (2)
               edge[E] node {} (0)
           (7) edge[E] node {} (6);
\end{tikzpicture}
\end{subfigure}
\hfill
\begin{subfigure}[b]{0.3\linewidth}
\centering
\begin{tikzpicture}
	\foreach \phi in {0,...,5}{
    	\node[vertex] (\phi) at (360/8 * \phi:2cm) {$w_\phi$};
     }
     \foreach \phi in {6,...,7}{
    	\node[vertex] (\phi) at (360/8 * \phi:2cm) {};
     }
     \draw (1) edge[SE] node {$+\Delta$} (0)
  			   edge[SN] node {$-\Delta$} (2)
               edge[E] node {} (6)
  		   (3) edge[SE] node {$+\Delta$} (2)
               edge[FE] node {} (0)
               edge[E] node {} (6)
               edge[SN] node {$-\Delta$} (4)
           (5) edge[SE] node {$+\Delta$} (4)
               edge[E] node {} (2)
               edge[SE] node {$-\Delta$} (0)
           (7) edge[E] node {} (6);
\end{tikzpicture}
\end{subfigure}
\hfill
\begin{subfigure}[b]{0.3\linewidth}
\centering
\begin{tikzpicture}
	\node[vertex] (0) at (360/8 * 0:2cm) {$w_0$};
    \node[vertex] (1) at (360/8 * 1:2cm) {$w_1$};
    \node[vertex] (2) at (360/8 * 2:2cm) {$w_{2,6}$};
    \node[vertex] (3) at (360/8 * 3:2cm) {$w_{3,7}$};
    \node[vertex] (4) at (360/8 * 4:2cm) {$w_4$};
    \node[vertex] (5) at (360/8 * 5:2cm) {$w_5$};
    \node[vertex] (6) at (360/8 * 6:2cm) {$w_8$};
    \node[vertex] (7) at (360/8 * 7:2cm) {$w_9$};

     \draw (1) edge[SE] node {$+\Delta$} (0)
  			   edge[SN] node {$-\Delta$} (2)
               edge[E] node {} (6)
  		   (3) edge[SE] node {$+2\Delta$} (2)
               edge[FE] node {} (0)
               edge[SE] node {$-\Delta$} (6)
               edge[SN] node {$-\Delta$} (4)
           (5) edge[SE] node {$+\Delta$} (4)
               edge[SE, pos = 0.6] node {$-\Delta$} (2)
               edge[FE] node {} (0)
           (7) edge[SE] node {$+\Delta$} (6)
           (7) edge[SN] node {$-\Delta$} (0);
\end{tikzpicture}
\end{subfigure}
\caption{\small{Sampled edge (black solid), sampled non-edge (black dashed), edge (gray solid) and fixed edge (gray dotted). If there are no fixed edges/non-edges, $W$ will have length 4 (left). Otherwise, longer sequences are required, as shown in the center and right examples. The right example demonstrates vertices may be revisited.}}
\label{samplingstep}
\end{figure*}
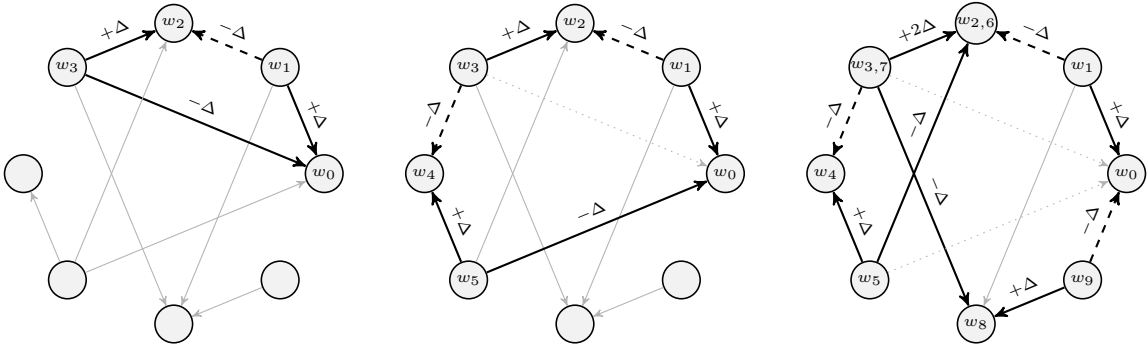

\begin{enumerate}
    \item Sample $W_{n+1} \sim U(N_G(W_{n}) \setminus \{ W_{n-1} \})$, unless the set from which to sample is empty, in which case return $K_{\text{id}}$.
    \item If $W_0 \in M_G(W_{n+1}) \setminus \{W_n \}$, then return $K_{[W]}$, where $W\colonequals W_0 ... W_{n+1}W_0$. Otherwise sample $W_{n+2} \sim U(M_G(W_{n+1}) \setminus \{W_{n}\})$, unless the set from which to sample is empty, in which case return $K_{\text{id}}$. Let $n = n + 2$.
\end{enumerate}
In practice, returning $K_{id}$ is rare and occurs primarily in cases of extreme sparsity; specifically when some vertex has only one in-neighbour. Figure \ref{samplingstep} graphically represents several example sequences $W$. 

\subsection{Conditional Distribution of \texorpdfstring{$\Delta$}{ph}}
\label{sec:conddist}

Suppose we sample a vertex sequence $w \in z$ using the strategy proposed in Section \ref{kss}. To ease notation, define $G_{low}$ and $G_{up}$ as the graphs obtained at $\Delta_{low}$ and $\Delta_{up}$ respectively. It is often not possible to sample $w$ from $G_{low}$, or to sample $w^r$ from $G_{up}$. This is why we collapse $w$ and $w^r$ into $z$; doing so ensures we can always sample $z$ from each graph in the support of $K_z(G,\cdot)$.

Suppose $(\Delta_{low}, \Delta_{up})$ is non-empty. 
Let $G^*$ be any graph obtained from sampling some $\Delta^* \in (\Delta_{low}, \Delta_{up})$. Define $\alpha_{int}$ as the probability of sampling $z$ from $G^*$. 
This quantity is the same regardless of the chosen $G^*$. 
If on the other hand the range $(\Delta_{low}, \Delta_{up})$ is empty, let $\alpha_{int}$ take an arbitrary finite value.

The random variable $\Delta$ then has conditional measure

\begin{equation} \label{condmeasure}
\frac{1}{d}\left( \mathbb{V}_{G_{low}}(z)\delta_{\Delta_{low}} + \mathbb{V}_{G_{up}}(z)\delta_{\Delta_{up}} + \alpha_{int} \sum_{\Delta_{low}<\tilde{\Delta}<\Delta_{up}} \delta_{\tilde{\Delta}} \right)
\end{equation}
where $d :=  \mathbb{V}_{G_{low}}(z) + \mathbb{V}_{G_{up}}(z) + \alpha_{int}\max(\Delta_{up}-\Delta_{low}-1,0)$.

$\mathbb{V}_{G_{low}}(z)$, $\mathbb{V}_{G_{up}}(z)$ and $\alpha_{int}$ are easily computed by following the details of the kernel selection strategy in Section \ref{kss}. 
Algorithm \ref{weightedsampler} gives pseudo-code for one iteration of the sampler. 

By construction, the chain is reversible with respect to the uniform distribution. Proposition \ref{allstatesweighted} holds by additionally showing the chain is connected.

\begin{proposition} \label{allstatesweighted}
The Markov chain is ergodic with respect to the uniform distribution on $\mathcal{G}$.
\end{proposition}

\begin{algorithm}[tbp]
\small
\caption{\small{One Iteration of the Weighted Graph Sampler (WGS)}}
\label{weightedsampler}
\begin{algorithmic}[1]
\Require $G$, $\mathcal{F}$
\State $\text{visits}(uv) \gets 0$; 
		$\text{edges} \gets \{\}$; 
        $(\Delta_{low}, \Delta_{up}) \gets (-\infty,\infty)$; 
        $W_{-1} \gets *$; 
        $n \gets 0$
\State $W_0 \sim U(\{ v \in V\colon N_G(v) \neq \emptyset \})$ 

\Repeat
\If{ $N_G(W_n) \setminus \{W_{n-1}\} =\emptyset$}
\Return $G$
\Else{} $ W_{n+1} \sim U(N_G(W_n) \setminus \{W_{n-1}\})$
\EndIf
\If{$W_0 \in M_G(W_{n+1}) \setminus \{W_n\}$} 
$W_{n+2} \gets W_0$
\ElsIf{$M_G(W_{n+1}) \setminus \{W_n\} = \emptyset$}
\Return $G$
\Else{}
$W_{n+2} \sim U(M_G(W_{n+1}) \setminus \{W_n\})$
\EndIf
\State $\text{edges} \gets \text{edges} \cup \{W_{n+1}W_n, W_{n+1}W_{n+2}\}$

\State $\text{visits}(W_{n+1}W_n) \gets \text{visits}(W_{n+1}W_n)+ 1$
\State$\text{visits}(W_{n+1}W_{n+2}) \gets \text{visits}(W_{n+1}W_{n+2}) - 1$
\State $n \gets n + 2$
\Until {$W_n = W_0$}
\ForAll{$\text{edge} \in \text{edges}$}
\If{$\text{visits}(\text{edge}) > 0$}
$\Delta_{low} \gets \max \left(\Delta_{low}, - \floor{\frac{c_G(\text{edge})}{\text{visits}(\text{edge})}}\right)$
\EndIf
\If{$\text{visits}(\text{edge}) < 0$}
$\Delta_{up} \gets \min \left(\Delta_{up}, \floor{\frac{c_G(\text{edge})}{\text{visits}(\text{edge})}}\right)$
\EndIf
\EndFor
\State Sample $\Delta$ according to \eqref{condmeasure}
\ForAll{$\text{edge} \in \text{edges}$}
\State $c_G(\text{edge}) \gets c_G(\text{edge}) + \text{visits}(\text{edge})\Delta$
\EndFor
\end{algorithmic}
\Return G
\end{algorithm}

\section{Simulation Study}
\label{simulations}
\begin{figure*}[tbp]
\centering
\includegraphics[width = \textwidth]{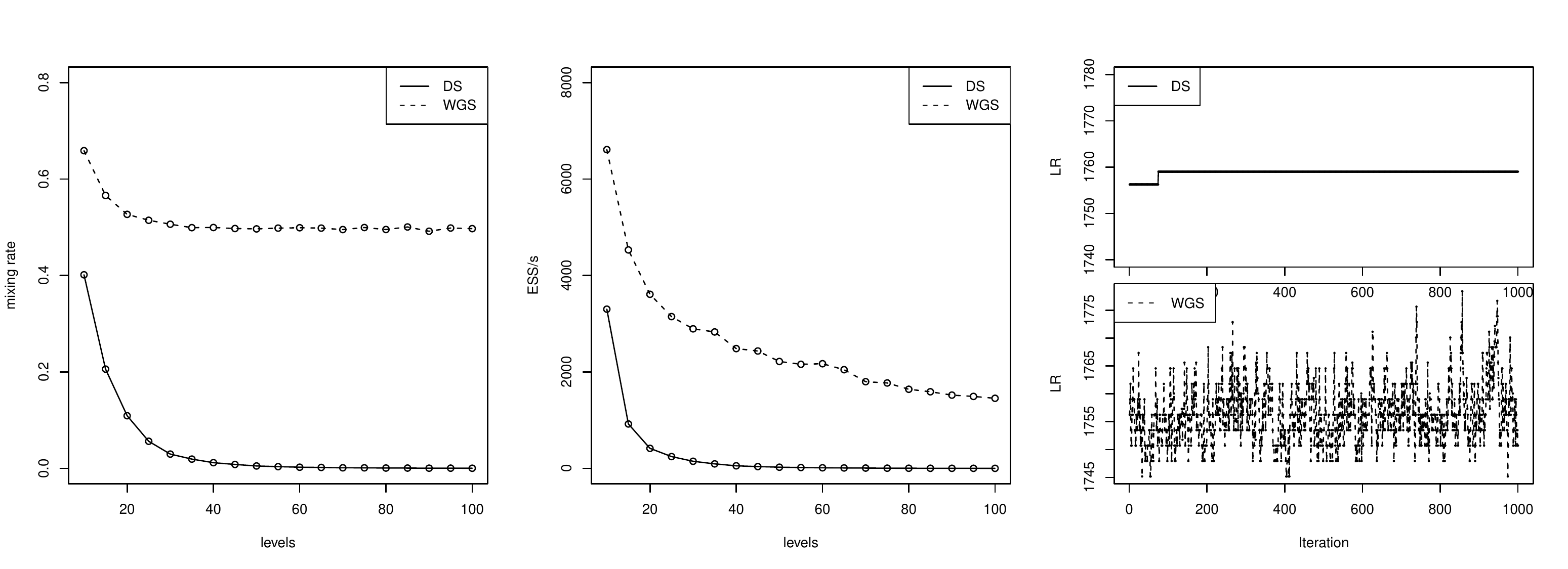}
\caption{{\small Comparative performance of WGS and DS in simulation study of Section \ref{simulations}.}}
\label{sparsityplots}
\end{figure*}

Methods used in this section, and in Section \ref{applications}, were programmed in C/C++, and run on an Intel Core i5-6360U 2GHz CPU.
Here we investigate the comparative performance of Algorithm \ref{weightedsampler} (WGS) and the Diaconis \& Sturmfels chain (DS) introduced in Section \ref{sdks} in sparse contingency tables.

Let $A$ and $B$ be independent, uniform categorical variables with $L$ levels. For each $L$ in $10,15, ... 100$, we simulate 200 bivariate samples and construct the corresponding $L \times L$ contingency table. We then use the samplers to approximately sample from the uniform distribution over all tables with the same margins, and obtain 100,000 samples of the likelihood ratio statistic computed from these tables. Thinning was set to $L$.

Define the (empirical) mixing rate of a Markov chain to be the observed proportion of iterations that change the state of the chain. Figure \ref{sparsityplots} shows plots this quantity for each sampler as we varied $L$. In addition Figure \ref{sparsityplots} plots the effective sample size per second (ESS/s) computed from the sample of LR statistics.

The mixing rate of DG rapidly approaches zero, while for WGS it levels at around 0.5. The effective sample size per second of DG rapidly falls rapidly, in line with the mixing rate, while it reduces at a much slower rate for WGS. When $L=100$, WGS appears to be over 1000 times more efficient than DG. Figure \ref{sparsityplots} provides trace plots of the likelihood ratio statistics.

\citet{eisinger_2017} develop efficient SIS methods for sampling tables from the uniform distribution over all tables with given margins. Their method labeled $SIS-G$ (coded in C) took 227 seconds to produce 1000 samples of $100 \times 100$ tables with both margins equal to $(5,1,...,1)$, evaluated on a laptop with a 2.2 GHz Intel Core i7 processor. The authors provide code for a cell-by-cell SIS method SIS-G*. Using SIS-G* on the same example, we estimated an ESS/s of around 7. Figure \ref{sparsityplots} shows that with $L=100$, the ESS/s for the LR statistics produced by WGS was over 1400. Therefore, it appears WGS can produce independent tables orders of magnitude faster than available SIS methods in large, spare tables.

\section{Applications}
\label{applications}
Reported standard errors were computed using spectral methods from R's coda package. These estimates were compared to those obtained using batch means, and where feasible, bootstrapping. These latter estimates are not reported as there was little discernible difference from those obtained by spectral methods. Thinning used in each method was set to approximately equate the resulting standard errors. We used burn-in equivalent to 20\% of samples obtained.

\subsection{Ecological Networks}
\label{ecologicalnetworks}

A food web encodes predator-prey relationships within a group of species, and has a natural representation as a digraph. Each species is a node in the graph and a link exists from species A to species B iff B consumes A. 

Ecologists wish to identify and explain structural patterns in observed food webs including motifs, diet contiguity, intervality, connectance and compartmentalization. We will focus on assessing the tendency towards compartmentalization in food webs. Compartmentalization describes the extent to which species can be partitioned into distinct groups such that linkage density \textit{within} groups is greater than that \textit{between} groups \citep{girvan_2002,krause_2003}. The level of compartmentalization in food webs is an important determinant of the spread of ecological perturbations. If a food web is highly compartmentalized, a perturbation should spread quicker within groups than between groups. Thus, higher compartmentalization should reduce systemic risks in the network and increase stability. Compartmentalization may also determine the effect of a perturbation to the network. An effect reducing compartmentalization, like the introduction of a generalist predator, may reduce the stability of the network \citep{guimera_2010}.

\begin{figure*}[tbp]
\centering
\includegraphics[width = 0.7\textwidth]{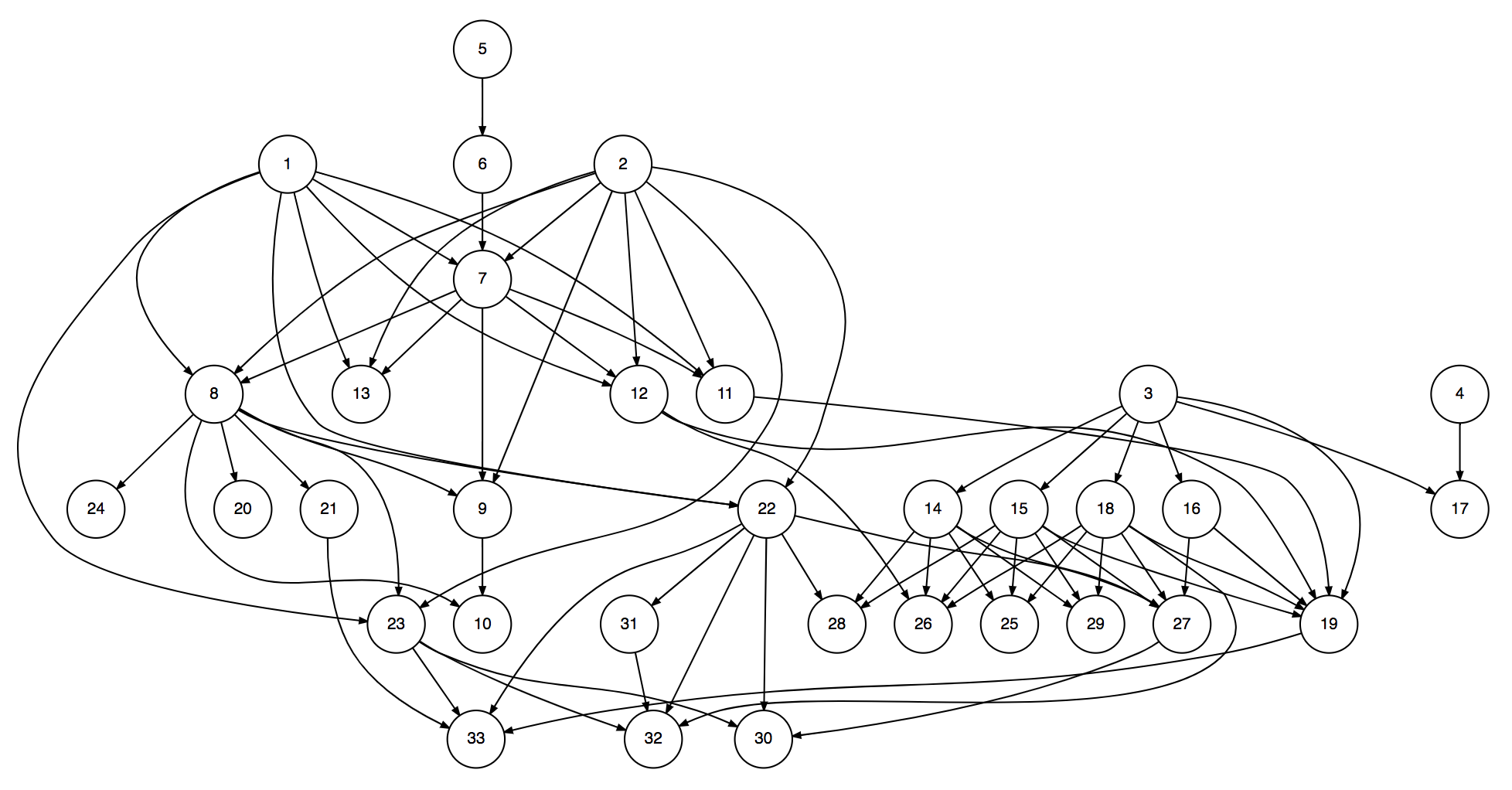}
\caption{{\small Food web of the Chesapeake bay ecosystem.}}
\label{chesapeakebay}
\end{figure*}

Figure \ref{chesapeakebay} depicts the food web of 33 species in the Chesapeake bay in the summer. The data was collected by \citet{baird_1989} and is provided in the R package Cheddar \citep{cheddar}. 

\citet{pimm_1980} proposed a statistic $\bar{C}$ to measure the level of compartmentalization in a food web. Here we describe a directed analogue of this statistic. Let $G$ represent a food web of $n$ species, and $i$ and $j$ be two species. Let $c_{ij}$ be the number of shared predators of species $i$ and $j$ as a proportion of the total number of predators of $i$ and $j$. $\bar{C}$ is then the mean of the off-diagonal elements of $(c_{ij})$.

$$
\bar{C} \colonequals \frac{1}{n(n-1)}\sum_{i=1}^{n} \sum_{j=1, j\neq i}^n c_{ij}
$$
$\bar{C}$ takes values in $[0,1]$ and higher values are associated with greater levels of compartmentalization.

We begin by testing whether the observed level of $\bar{C}_0 = 0.0260$ can be considered high when compared to the set of all graphs with the same in-degree and out-degree sequence as $G$ \citep{ulrich_2007}. With thinning of 5, Algorithm \ref{binarysampler} (UGS) took around 2 second to obtain 100,000 samples. The estimated p-value was $0.0163 \pm 4.3\mathrm{e}{-4}$, complementing previous results suggesting food webs have high compartmentalization when compared to random graphs where species have an equal probability of linking to each other species \citep{krause_2003,rezende_2009}.

\citet{chen_2007} develop a sequential importance sampling strategy SIS\_CP1 for sampling zero-one tables, but which is equivalently capable of uniformly sampling unweighted digraphs with fixed degree sequence and an arbitrary pattern of structural non-edges. SIS\_CP1 took 33 seconds to obtain 100,000 samples, estimating a p-value of $0.0158 \pm 4.3\mathrm{e}{-4}$.

\citet{guimera_2010} find that compartmentalization observed in real food webs is not unusual when compared to networks generated under niche models, and conclude that `compartmentalization can be explained solely by the niche-valued ranking of species'. 

We attempt to test this hypothesis for the Chesapeake bay food web. We compute the chain averaged trophic level \citep{williams_2004} for each species, and assume any given species is forbidden from consuming other species with a higher trophic level. The resulting forbidden links should help to control for the food web's trophic structure. The assumption induces 565 forbidden edges in the null distribution.

Again using thinning of 5, UGS took 2 seconds to obtain 100,000 samples. The estimated effective sample size was over 95,000, giving an estimated p-value of $0.0568 \pm 7.5\mathrm{e}{-4}$. At a significance level of $\alpha = 0.05$, we can no longer conclude that the level of compartmentalization in the Chesapeake food web is unusual under the null distribution. Our method of determining trophic structure is relatively crude, and a closer analysis of the food web is warranted before drawing any conclusions.

SIS\_CP1, on the other hand, took 24 seconds to run and over 97\% of the samples produced were discarded as invalid, leaving only 3,069 to be used for estimation. The estimated p-value was $0.0558 \pm 6.3e\mathrm{-3}$. Using alternative methods to calculate the species' trophic levels gives rise to other sets of forbidden edges. For some such patterns, SIS\_CP1 was unable to construct a single valid sample. SIS\_CP1 cannot reliably sample graphs in the face of arbitrary sets of fixed non-edges.

Forbidden links have a long history in ecological networks, and result from factors including spatio-temporal uncoupling and morphological and physiological-biochemical constraints \citep[see][]{olesen_2010}. A realistic specification of the null distribution of a hypothesis test must take into account forbidden links, particularly as they can change the outcome of the test. Our algorithm provides this flexibility, even in large sparse graphs. Other MCMC methods generally require a Markov basis, which is computationally infeasible for networks with more than a few nodes.

\subsection{Incomplete Tables}
\label{incomplete}

A sample of size $N$ consisting of measurements on two categorical variables can be represented as a two-way contingency table of dimensions $I \times J$, where $I$ and $J$ represent the number of levels of the first and second variables respectively.

A table with structurally fixed cells is referred to as incomplete. Incomplete tables arise in several contexts. Particular combinations of the variables may be impossible, forcing zero entries in the corresponding cells. Alternatively, some observations may be missing. In some contexts, researchers may wish to fit composite models by partitioning the cells into subsets, and fitting a separate log-linear model for each group \citep{goodman_1963, goodman_1968, fienberg_1969}). See \citet{bishop_1969} for extensive examples of incomplete tables.

\citet{pearson_1904} introduce a mobility table recording the occupations of 775 fathers and their sons. There are 14 separate job classifications. Given a table $t$ and a log-linear model, define the chi-squared statistic $\chi^2$ as

$$
\chi^2 = \sum_{i=1}^r \sum_{j=1}^{c} \frac{(t_{ij} - \hat{m}_{ij})^2}{\hat{m}_{ij}}
$$
where $\hat{m}$ is a matrix representing the MLEs of expected cell counts under the log-linear model. For the independence model, we observe $\chi^2 = 1005.45$. This appears high, indicating that the model of independence is not appropriate.

\citet{goodman_1965} propose that sons are liable to inherit their occupational status from their fathers, but conditional on them changing occupation, their occupation choice is considered independent from their father's status. This resulting model of `quasi-perfect mobility' fixes the diagonals of the table. 

Alternatively, \citet{fienberg_1969} propose a graphical procedure to determine cells with large interactions. They identify 14 such cells, namely $(1,1)$, $(13,13)$, $(11,11)$, $(7,1)$, $(3,3)$, $(3,1)$, $(7,7)$, $(2,1)$, $(2,2)$, $(4,4)$, $(12,12)$, $(10,10)$, $(6,8)$ and $(6,12)$. 10 of these are along the diagonal. 

We wish to test whether the remaining observed counts are plausible under the independence model, \textit{conditional} on the 14 cells being fixed. The model of independence applied to tables with structural zeros is known as the quasi-independence model \citep{goodman_1968}. The MLEs of the expected cell counts can easily be computed using the IPFP procedure.

\citet{diaconis1985} propose the uniform distribution on the set of all $I \times J$ tables with the same margins as an alternative to the independence hypothesis. This can also be applied to incomplete tables. With obvious modifications, our Markov chain can also be used to sample from the hyper-geometric distribution to evaluate the hypothesis of quasi-independence.

Fixing the aforementioned 14 cells and applying the quasi-independence model yields $\chi^2 = 345.21$. Using thinning of 50, WGS obtained 100,000 samples in 3.5 seconds. The estimated p-value was $0.99987 \pm 4.1e-5$ and the estimated effective sample size was over 70,000. Conditional on fixing the 14 cells, the table appears to fit the quasi-independence hypothesis extremely well. The oddly high p-value may perhaps be less surprising if you consider that Feinberg's graphical procedure has chosen the pattern of fixed cells in order to remove large deviations from independence.

\citet{diaconis_sturmfels_1998} propose an alternative Markov chain based on computational algebra, and requiring the computation of a Markov basis. Using the software Macaulay 2 (\citet{M2}), we were unable to compute a Markov basis for the support of the distribution.

\citet{chen_2007} propose an SIS algorithm for performing the conditional volume test on incomplete tables. They provide an implementation capable of handling at most one structural zero in each column. Therefore, we can use their method to test the quasi-perfect mobility model. The algorithm produced 100,000 samples in 1 second, however the variation in the importance weights was extremely high, indicating that the sampling distribution is far from uniform and giving a negligible effective sample size.

It appears current SIS methods for incomplete tables are not scalable in the size of the table. This is also true for Markov chain methods using computational algebra. Our sampler appears to largely overcome these difficulties.

\section{Discussion}
\label{discussion}

This paper has developed a new class of MCMC samplers for two important sampling problems. First, for sampling from the set of unweighted graphs respecting prescribed vertex  degrees. Second, for sampling from the set of weighted graphs respecting prescribed vertex strengths.

The samplers appear more efficient than existing methods in sparse settings, and also when there are arbitrary observed edges/non-edges. We have presented examples where alternative MCMC methods are infeasible as they rely on computing a Markov basis, and where existing SIS methods perform poorly. In contrast, our methods do not require computing a Markov basis, and appear to be orders of magnitude more efficient in these examples.

State-dependent mixing of Markov kernels is a general concept, and the specific implementation of our samplers is not unique. The technique could be used to develop alternative samplers specialized to particular settings. The methods can be readily extended to sample from arbitrary distributions known up to a normalization constant. Thus, the samplers can be adapted to carry out Bayesian network tomography in the case of a star network topology. In contrast, SIS methods are not readily adaptable to more general distributions. A theoretical analysis of the mixing times of the new samplers is beyond the scope of this paper. Future work could try to establish whether the chains are rapidly mixing.

\section{Acknowledgements}

The work of the first author supported by an EPSRC Research Studentship.

\bibliography{jascott}

\appendix

\section{Proofs}

\begin{proof}[\bf{Proof of Lemma \ref{symmetrictheorem}}]
The identity decomposition immediately shows that the `only if' part holds. For the `if' part, we must show reversibility of $Q$ with respect to $\pi$. Without loss of generality (and for notational simplicity) assume $\pi$ and $\mathbb{V}$ are dominated by one-dimensional Lebesgue measure. Reversibility is then defined as 

$$
\int_A\pi(x)Q(x,B )\diff x = \int_B \pi(x)Q(x,A)\diff x \quad \text{ for all } A,B \in \mathcal{\mathcal{B}} .
$$

Fix any $A$ and $B$ in $\mathcal{B}$, and define the densities $\{f_x \}$ as in Definition \ref{symmetrickernel}. Then

\begin{equation*}
\begin{split}
\int_A\pi(x)Q(x,B )\diff x &= \int_A \int_\mathcal{Z} \int_{B} \pi(x)K_z(x,y)f_x(z) \diff y \diff z \diff x\\
&= \int_A \int_\mathcal{Z} \int_{B} \pi(y)K_z(y,x)f_y(z) \diff y \diff z \diff x
= \int_B\pi(x)Q(x,A )\diff x
\end{split}
\end{equation*}
as required. In the first step we have expressed the integral using densities. In the second, we use reversibility of each $K_z$ with respect to $\pi$, and the fact that for each $z \in \mathcal{Z}$ and $x \in \mathcal{X}$, $f_x(z) = f_y(z)$ for $K_z(x,\cdot)$- a.e. $y \in B$. A simple change of variables then yields the result.
\end{proof}

\begin{proof}[\bf{Proof of Proposition \ref{finitez}}]

Fix $G$ in $\mathcal{G}$. Define the second-order Markov chain $(Y_n)_{n\geq 0}$, where $Y_n \colonequals (W_{n-1}, W_n, G_n)$ and $G_n$ is defined as follows. Let $G_0 = G$, otherwise if $n$ is odd, let $G_n$ be the graph obtained after $W_nW_{n-1}$ is removed from $E(G_{n-1})$. If $n$ is even, $G_n$ is the graph obtained after $W_{n-1}W_n$ is added to $E(G_{n-1})$. Define $\mathcal{Y}$ as the set of points reachable from $(*, W_0, G)$ for some $W_0$ in $\{v : N_G(v) \neq \emptyset \}$. Let $D \colonequals (\mathcal{Y},E)$ be the digraph underlying this chain and define $A$ as the subset of points $(v,u,G')$ in $\mathcal{Y}$ for which $G' \in \mathcal{G}$. Let $T \colonequals  \inf \{n \geq 1 \colon Y_n \in A\}$ be the first passage time of $A$. Proposition \ref{finitez} is equivalent to showing $\mathbb{E}(T) < \infty$.
The following holds true, and will be shown at the end of this proof.
\begin{equation}
\label{altaccess}
\text{From any }(v,u,G')\in\mathcal{Y}\text{, there exists a simple path to }A.
\end{equation}

We can bound the probability of traversing each edge in $D$ from below by some constant $p>0$. Let $N$ denote the size of $\mathcal{Y}$. Suppose the chain is at some state $y \notin A$. By \eqref{altaccess}, this implies the probability of hitting $A$ within the next $N$ steps is bounded from below by $p^{N}$.
Hence,
\begin{equation*}
\begin{split}
\mathbb{E}(T) & = \sum_{n=1}^{\infty}nP(T=n) \leq N\sum_{k=1}^{\infty}kP(k-1 < T/N \leq k)
 \leq N\sum_{k=1}^{\infty} kp^N(1-p^N)^{k-1} = Np^{-N}<\infty.
\end{split}
\end{equation*}

Thus it remains to show \eqref{altaccess}. We need the following observation repeatedly:  for any 
$(v_1, u_1, G_1)(v_2,u_2,G_2) \in E(D)$:
\begin{equation}\label{altedgecondition}
(v^*,u_2,G_2)(u_2,u_{1},G_{1}) \in E(D)\text{ if and only if }(v^*,u_2,G_2) \in \mathcal{Y}\text{ and }v^* \neq v_2. 
\end{equation}
We now show \eqref{altaccess}. By definition, there exists a point $y_0 := (*, u_0, G)$ and a walk $y_0 ... y_k$ in $D$ such that $y_k = (v,u,G')$. Given that $V(D)$ is finite, continuing an arbitrary walk along $D$ from $y_k$ implies we must eventually either return to $A$, or visit a graph already seen along the walk. Denote the vertex visited at the $l$th step of this walk by $y_l = (u_{l-1}, u_l, G_l)$. If we revisit $A$ we are done, otherwise define
$$
n := \inf \{l >k : G_l = G_m \text{ for some } m < l\}.
$$
The condition $G_n = G_m$ implies that $u_n = u_m$. Additionally $u_{n-1} \neq u_{m-1}$, otherwise this would imply $G_{n-1} = G_{m-1}$, which contradicts the definition of $n$. By \eqref{altedgecondition}, $(u_{m-2},u_{m-1},G_{m-1})(u_{m-1},u_{m},G_{m}) \in E(D)$ implies that $(u_{n-1}, u_n, G_n)(u_m,W_{m-1}, G_{m-1}) \in E(D)$. Thus we can traverse to $(u_m,u_{m-1}, G_{m-1})$. Iteratively applying \eqref{altedgecondition} (which we can do as $u_{l+1} \neq u_{l-1}$ for all $l\geq 0$) implies we can reach a state with graph $G = G_0$, which must be in $A$, completing the proof of \eqref{altaccess}.
\end{proof}

\begin{proof}[\bf{Proof of Lemma \ref{symdecomp}}]
We first show that $(K,\mathbb{V})$ is a symmetric decomposition. Fix any $G \in \mathcal{G}$ and any $z \in \mathcal{Z}$, and let $w$ be a representative of $z$. Let $p$ refer to the statement `$\mathbb{V}_G(z) = \mathbb{V}_{G^*}(z)$ for all $G^*$ for which $K_z(G,G^*)>0$'. It suffices to show that $p$ is true.

Consider a Markov chain with kernel $K_z$ and current state $G$. Suppose the chain remains unchanged after one iteration of Algorithm \ref{binarysampler}. Then $p$ is true trivially. Without loss of generality, suppose the swaps corresponding to $w$ are viable, and the chain moves to some $G^* \in \mathcal{G}$. Remark \ref{welldefined} implies swaps corresponding to $w^r$ are not viable. Since the swaps corresponding to a sampled vertex sequence must be viable, $\mathbb{V}_G(z)$ is the probability $W = w$ given the chain is at $G$. 

At $G^*$, the swaps corresponding to $w^r$ are viable. By an analogous argument, it follows that $\mathbb{V}_{G^*}(z)$ is the probability $W = w^r$ given the chain is at $G^*$. One can deduce from Algorithm \ref{binarysampler} that the probability $W = w$ given the chain is at $G$ is equal to the probability $W = w^r$ given the chain is at $G^*$. This holds because the degree sequence is the same for either state.

We now show that each $K_z \in K$ is reversible with respect to the uniform distribution. This is implied by detailed balance. Specifically, for each $K_z \in K$ we show 

\begin{equation*}
K_z(G,G^*) = K_z(G^*,G) \quad \text{ for all } G, G^* \in \mathcal{G}.
\end{equation*}

Fix any $G$ and $G^*$. $K_z(G,G^*) = 1$ if and only if $K_z(G^*,G) = 1$, because applying two iterations of a Markov chain with kernel $K_z$ from some current state $G'$, returns $G'$. The result follows by additionally observing that $K_z(G,G^*)$ can only be zero or one.
\end{proof}

\begin{proof}[\bf{Proof of Proposition \ref{allstates}}]

The proposition is implied by Lemma \ref{symdecomp}, and additionally showing the chain is connected. 

Fix any $G,G' \in \mathcal{G}$, and suppose the current state of the chain is $G$. Form a digraph $H$ as follows. For each vertex pair $uv$, if $uv \in E(G)$ and $uv \notin E(G')$, add a \textit{red} edge $uv$ to $E(H)$. If $uv \notin E(G)$ and $uv \in E(G')$, add a \textit{blue} edge $uv$ to $E(H)$. Define an \textit{alternating} cycle as a cycle whose edges are alternately red and blue. $G$ and $G'$ are equivalent if and only if $H$ has no edges.

Then 
$H$ is the union of a finite sequence of edge-disjoint alternating cycles.

Fix any alternating cycle $v_0v_1...v_k v_0$ implied by this, ordered so that the $v_0v_1$ is red. The Markov chain can sample $W := v_0v_1...v_k v_0$ with positive probability, yielding a new graph $G''$, whilst removing all edges in $H$ corresponding to this cycle. Iterate until $H$ has no more edges.
\end{proof}
\begin{proof}[\bf{Proof of Proposition \ref{stronglyconnected}}]

For a given $\mathcal{F}$, the map from $\mathcal{G}$ to $B_{\mathcal{G}}$ is injective, so the sampler can be thought of as a Markov chain ergodic with respect to the uniform distribution on $B_{\mathcal{G}}$. 

We briefly describe how to view the Markov chain as operating on $B_\mathcal{G}$. An initial vertex $v_j$ is sampled from $V$. The chain now samples $u_i$ from the out-neighborhood of $v_j$ and replaces the edge $v_ju_i$ with $u_iv_j$. If $G$ is undirected, additionally switch $v_iu_j$ with $u_jv_i$. Continue walking along the vertices of the graph in this manner until the sampler returns to the initial vertex for the first time.

Without loss of generality, suppose $G$ is directed. Fix $u_i \in S_k$ and $v_j \in S_l$. If no edge is incident to $u_i$ and $v_j$ then $ij \in \mathcal{F} \subseteq \tilde{\mathcal{F}}$. Otherwise if $k \neq l$, edges between $S_k$ and $S_l$ are uniformly in one direction; say from $S_k$ to $S_l$. Suppose the Markov chain on $B$ traverses $u_iv_j$, replacing it with $v_ju_i$. Returning to the initial vertex requires traversal of $v_ju_i$. Hence, $u_iv_j$ can be flipped only an even number of times, and the direction is unchanged. By Lemma \ref{allstates}, $ij\in \tilde{\mathcal{F}}$. If $k=l$, $u_iv_j$ can be switched odd number of times, so $ij \in \tilde{\mathcal{F}}$. The undirected case holds by an analogous argument.
\end{proof}


\begin{proof}[\bf{Proof of Proposition \ref{allstatesweighted}}]
It suffices to show connectedness. Define a metric $d: \mathcal{G} \times \mathcal{G} \rightarrow \mathbb{N}_0$ on $\mathcal{G}$ by

$$
d(G,G') \colonequals \sum_{u \in V}\sum_{v \in V} |c_G(uv) - c_{G'}(uv)|
\quad \text{ for all } G, G' \in \mathcal{G}.
$$
Then $(\mathcal{G},d)$ is a metric space. Fix any two distinct graphs $G,G'\in \mathcal{G}$, and suppose the current state of the chain is $G$. It suffices to show that one can construct a sampling step yielding a new graph \textit{strictly} closer to $G'$ in this metric space.

Let $n_{uv} := c_G(uv) - c_{G'}(uv)$ for each vertex pair $uv$. We form a multi-graph $H$ as follows. If $n_{uv}>0$, add $n_{uv}$ \textit{red} copies of the \textit{direction reversed} edge $vu$ to $E(H)$, while if $n_{uv} < 0$, add $-n_{uv}$ \textit{blue} copies of $uv$ to $E(H)$. The graphs $G$ and $G'$ are equivalent if and only if $H$ has no edges. Define an \textit{alternating} cycle in $H$ as a cycle whose edges are alternately $red$ and $blue$. 

It can be shown that 
$H$ can be expressed as the union of a finite number of edge-disjoint alternating cycles.

Fix any such  alternating cycle $v_0v_1...v_lv_0$ in $H$. Order the cycle so that $v_0v_1$ is red. Letting $\mathbb{O}_n$ denote the set of odd natural numbers less than or equal to $n$, we define 

$$k := \inf \{ n \in \mathbb{O}_{l-2} : v_nv_{n+1} \notin \mathcal{F} \}$$
where we let $\inf \emptyset := l$. 

Under Algorithm \ref{weightedsampler}, there is a positive probability of sampling the vertex sequence $v_0v_1...v_kv_0$ given the chain is at $G$. Sampling $\Delta = -1$ along this vertex sequence returns a new graph $G''$, removing at least three edges from $H$ whilst adding at most one. Hence

$$
d(G'',G') \leq d(G,G') - 2.
$$
\end{proof}
\end{document}